\documentclass[pre,showpacs,showkeys,superscriptaddress,nofootinbib,floatfix,onecolumn,amsmath]{revtex4}
\usepackage{color}
\usepackage{amsfonts}
\usepackage{amsbsy}
\usepackage{mathrsfs}
\usepackage{graphicx}
\usepackage{subfigure}
\newcommand{\eeq}{\end{equation}}
\newcommand{\bea}{\begin{eqnarray}}
\newcommand{\eea}{\end{eqnarray}}

\def\lsim{\mathrel{\rlap{
\lower4pt\hbox{\hskip-3pt$\sim$}}
    \raise1pt\hbox{$<$}}}     %less than approx. symbol
\def\gsim{\mathrel{\rlap{
\lower4pt\hbox{\hskip-3pt$\sim$}}
    \raise1pt\hbox{$>$}}}     %greater than or approx. symbol

\begin{document}

\title{Thermodynamic properties of the finite--temperature electron gas 
\\ by the fermionic path integral Monte Carlo method}
\author{V.S.~Filinov}
\affiliation{Joint Institute for High Temperatures, Russian Academy of Sciences,
Izhorskaya 13 Bldg 2, Moscow 125412, Russia}
\author{P.R.~Levashov}
\affiliation{Joint Institute for High Temperatures, Russian Academy of Sciences,
Izhorskaya 13 Bldg 2, Moscow 125412, Russia}
\affiliation{Moscow Institute of Physics and Technology, 9 Institutskiy per., Dolgoprudny, Moscow Region, 141700, Russia}
\author{A.S.~Larkin}
\affiliation{Joint Institute for High Temperatures, Russian Academy of Sciences,
Izhorskaya 13 Bldg 2, Moscow 125412, Russia}
%\author{M.~Bonitz}
%\author{Zh. Moldabekov}
%\affiliation{Institut f\"ur Theoretische Physik und Astrophysik, Christian-Albrechts-Universit{\"a}t zu Kiel,
%Leibnizstrasse 15, 24098 Kiel, Germany}

\begin{abstract} 
The new {\em ab initio} quantum path integral Monte Carlo  approach has been developed and applied for the entropy 
difference calculations for the strongly coupled degenerated uniform electron gas (UEG), a well--known model of 
simple metals. Calculations have been carried out at finite temperature in canonical ensemble over the wide density and temperature ranges. 
Obtained data  may be crucial for density functional theory. 
Improvements of the developed approach include the Coulomb and exchange interaction of fermions in the basic Monte Carlo cell 
and its periodic images and the proper change of variables in the path integral measure.
The developed approach shows good agreement with available results for fermions even at temperature four times less 
than the Fermi energy and practically doesn't suffer from the ``fermionic sign problem'', which takes place in standard path 
integral Monte Carlo simulations of degenerate fermionic systems.   
Presented results include pair distribution functions, isochors and isotherms of pressure, internal energy  
and entropy change in strongly coupled and degenerate UEG in a wide range of density and temperature. 
\end{abstract}
\maketitle
%71.15.Nc 	Total energy and cohesive energy calculations
%05.30.Fk 	Fermion systems and electron gas
%05.70.Ce 	Thermodynamic functions and equations of state
%

\section{Introduction}
Entropy is a fundamental quantity in physics and is recognized as one of the most important and enigmatic 
concepts in nature \cite{Martin:entropy:2011}. In the studies of phase stability, phase transitions and reaction directions 
the free energies and entropies corresponding to the desired phases or reaction states are the most properties of interest. 
In this paper we applied the concepts of entropy to consideration of the one-component plasma (OCP), 
being a fundamental many-body model consisted of a single specie of charged particles, immersed in a rigid neutralizing background.
For electrons, OCP is often called ``uniform electron gas'' (UEG) or ``jellium'' and can be used
as a model of simple metals. Even though UEG itself does not represent a real physical system, its accurate description is
crucial for density functional theory (DFT)  \cite{kohn1965self,hohenberg1964inhomogeneous,filinov2020uniform}. 

Accurate data for the electron gas of the highly excited systems {\em at finite temperatures}, 
such as warm dense matter have been obtained by the path integral Monte Carlo (PIMC) techniques 
%e.g. {\em ab initio} simulations  
\cite{feynmanquantum,binder1996monte,ceperley1995path,zamalin1977monte,filinov2006introduction,egger1999crossover, filinov2001wigner}  
and for dense plasma in \cite{militzer2000variational,militzer2006first,filinov2000high} .
These results have been used for DFT calculations 
%\cite{pribram2015dft} 
as a basis for the local (spin-)density approximation (L(S)DA) and  more sophisticated gradient approximations  
\cite{karasiev2016importance,ramakrishna2020influence,karasiev2019exchange,
karasiev2014accurate,karasiev2019status,dornheim2018uniform,perdew1996generalized,perdew1996generalized1}. 

The cornerstone difficulty in the path integal Monte Carlo (PIMC) simulations of quantum
fermions is the ``fermionic sign problem''. 
Reliable Monte Carlo simulations at finite temperature in wide fermion density range have been carried out by a permutation 
blocking (PB) PIMC approach \cite{njp,prl2} and 
the configurational PIMC approach   (CPIMC)   \cite{cpp,prl1,schoof2011configuration,schoof2015towards,groth2016ab,dornheim2015permutation,malone2015interaction,malone2016accurate,blunt2014density}. 
The Dornheim, Groth and co-workers \cite{dornheim2015permutation,dornheim2016ab}
have developed an formalism, which allows to approach the thermodynamic limit 
and construct a highly accurate parametrization of the UEG exchange-correlation free energy.  
On the other habd the approach proposed in \cite{larkin2017pauli,larkin2017peculiarities} is based on the Wigner formulation of quantum 
mechanics. This approach has been realized the Pauli blocking of fermions without antisymmetrization of matrix elements as 
it used the derived effective pair pseudopotential in phase space 
preventing occupation of a quantum phase space cell by the two identical fermions.  

However data for entropy of the strongly coupled UEG are still unknown.  
Widely used approaches to calculate the entropy in canonical ensemble at finite temperature for arbitrary system of particles 
are based on the thermodynamic integration 
% [13]
\cite{frenkel2001understanding} 
and nonequilibrium work relations such as the Jarzynski equality 
%[14] 
\cite{jarzynski1997nonequilibrium} for the determination the changes in the free-energy 
and allowing to obtain the entropy difference. 
The microcanonical equivalent of the Jarzynski equality 
%[15] 
\cite{adib2005entropy} was used to calculate 
isoenergetic entropy differences using a switching parameter. 
The Wang-Landau techniques 
%[16,17,18] 
\cite{wang2001determining,wang2001efficient,shell2002generalization,davis2011calculation}
for the determination of entropy 
%, originally developed for discrete models but soon generalized to the continuous case [18]. WL techniques 
consist of performing a random walk in energy space to achiev a flat histogram of energies.
%The density of states is updated in the process, and finally converges to its true value.  

The motivation of the present work is to perform 
{\em ab initio} direct fermionic PIMC (FPIMC) simulations to calculate entropy at finite temperature in canonical ensemble. 
We modify the direct PIMC approach, that previously has been successfully applied to dense hydrogen,
hydrogen-helium mixtures, electron-hole plasmas in semiconductors and nonideal quark--gluon plasma \cite{EbelForFil}.  
To reduce the finite size effects of FPIMC and the influence of artificial periodic boundary conditions in the 
interparticle interaction we use the scheme proposed by Yakub et al. \cite{yakub2003efficient}. 
Here we have developed the same ideas to describe exchange interaction between fermions.
The numerical results presented in this article demonstrate the significant reducing of the 
``fermionic sign problem''. 
In section 2 we describe the basic concepts of the UEG model, its path integral description and 
mathematical formalism for entropy change calculations.  
In section 3 we present obtaines results for pair distribution functions, pressure and internal energy on the isochors and isotherms 
and the entropy changes in the strongly coupled and degenerate UEG in a wide range of density and temperature. 
Developed approach reveals good reliability  for temperatures in several times less than the Fermi energy. 
In section 4 we present some concluding remarks and possible perspectives.   
 
\section{Fermionic path integral Monte Carlo simulations}\label{theory}  
\subsection{Jellium model}

UEG, or jellium, is a quantum mechanical model of interacting electrons on a rigid neutralizing background. 
This model allows one to treat the quantum effects of electrons and their repulsive interaction rigorously. 
In this article we are going to simulate thermodynamic properties of strongly coupled UEG. 
Due to the strong electron interaction and absence of small physical parameters we can not use analytical methods based 
on different perturbation theories and have to deal with the density matrix of the system. Our efforts in this direction have resulted in the development of a new ``ab initio'' quantum fermionic path integral Monte Carlo approach (FPIMC)   
\cite{filinov2020uniform}. The UEG Hamiltonian contains the electron kinetic ${\hat K}$ and Coulomb energy 
 ${\hat  U}^c_{ee}$ contributions, the interaction of electrons with the neutralizing background ($\hat U^c_{ep}$) and the self-interaction energy 
 of the background ($\hat U^c_{pp}$). The rigid neutralizing background is simulated as an ideal gas of uncorrelated 
classical positive charges uniformly distributed in space. The path integral representation of density matrix is used to obtain the UEG partition function and  thermodynamic properties. A number of modifications to the standard PIMC scheme discussed below allowed us to significantly reduce the influence of the notorious ``fermionic sign problem''.
%Thermodynamic properties in FPIMC simulations are calculated by using estimators of thermodynamic values accounting for electron interaction 
%with the uncorrelated positive charges and interaction of positive uncorrelated charges. 
%Used estimators have to account for also the electron interaction itself.  
% ${\hat U}_{pp}=0$
%\textbf{\color{red} Above is the new text. Here should be some text about positive charges and background.}

Let us consider a neutral two-component Coulomb system of quantum electrons and classical positive charges in equilibrium 
with the general Hamiltonian, ${\hat H}={\hat K}+{\hat U}^c$, containing kinetic energy ${\hat K}$ and Coulomb interaction energy contributions,
${\hat U}^c = {\hat U}_{pp}^c + {\hat  U}^c_{ee} + {U}^c_{ep}$. 
%Discussed above modification in matrix elements of the density matrix 
%and estimators will be accounted for in the final formulas. 

The thermodynamic properties in the canonical ensemble with a given temperature $T$ and fixed volume $V$ are fully described 
by the density operator ${\hat \rho} = e^{-\beta {\hat H}}$, with the partition function 
\begin{equation}\label{q-def}
Z(N_e,N_p,V;\beta) = \frac{1}{N_e!N_p!} \sum_{\sigma}\int\limits_V
\rm dx \,\rho(x, \sigma ;\beta),
\end{equation}
where $\beta=1/k_BT$, and $\rho(q, \sigma ;\beta)$ denotes the diagonal elements of the density matrix in the coordinate representation 
at a given value $\sigma$ of the total spin. In Eq.~(\ref{q-def}), $x=\{x_e,x_p\}$ and $\sigma=\{\sigma_e\}$ are the spatial coordinates 
(in units of thermal wave lengths) of electrons and positive charges and spin degrees of freedom
of the electrons, i.e. $x_a=\{x_{1,a}\ldots x_{l,a}\ldots x_{N_a,a}\}$
and $\sigma_e=\{\sigma_{1,e}\ldots \sigma_{l,e}\ldots \sigma_{N_e,e}\}$.
The exact density matrix of interacting quantum systems is not known (particularly at low temperatures and high
densities), but can be constructed using a path integral representation~\cite{feynmanquantum} based on the operator identity,
\begin{equation}
e^{-\beta {\hat H}}= e^{-\epsilon {\hat H}}\cdot
e^{-\epsilon {\hat H}}\dots  e^{-\epsilon {\hat H}}, \quad \epsilon = \beta/(M+1)
\end{equation}
that involves $M+1$ identical high-temperature factors with a temperature $(M+1)T$, which allows us to 
rewrite the integral in Eq.~(\ref{q-def}) as 
\begin{eqnarray}&&\sum_{\sigma} \int\limits \rm dx^{(0)}\,\rho(x^{(0)},\sigma;\beta) =
%\nonumber\\&&
\int\limits  \rm dx^{(0)} \dots \rm  dx^{(M)} \, \rho^{(1)}\, \dots\rho^{(m)} \, \dots \rho^{(M-1)} \times
%\nonumber\\
\nonumber\\
&&\sum_{\sigma}\sum_{P_e} (-1)^{\kappa_{P_e}} \,
{\cal S}(\sigma, {\hat P_e} \sigma_{a}^\prime)\, \times
%\nonumber\\&&
{\hat P_e} \rho^{(M)}\big|_{x^{(M+1)}= x^{(0)}, \sigma'=\sigma},
\label{rho-pimc}
\end{eqnarray}
where index $m=0, \dots, M$ labels the off--diagonal high--temperature density matrices
$\rho^{(m)}\equiv \rho\left(x^{(m)},x^{(m+1)};\epsilon \right) =
\langle x^{(m)}|e^{-\epsilon {\hat H}}|x^{(m+1)}\rangle$. 
The spin gives rise to the spin part of the density matrix (${\cal S}$) with exchange effects accounted for by the permutation
operator  $\hat P_e$ acting on the electron coordinates $q^{(M+1)}$ and spin projections $\sigma'$. The
sum is over all permutations with parity $\kappa_{P_e}$. 

Let us consider the off-diagonal elements of the density matrix $\rho(\tilde{x}, \sigma ; x^{\prime}, \sigma^{\prime} ; \beta)$.  
With the error of order $1/M^2$,
arising from neglecting the commutator $\epsilon^2/2 \left[K,U^c\right]$, each high-temperature factor can
be presented in the form
$\langle x^{(m)}|e^{-\epsilon {\hat H}}|x^{(m+1)}\rangle \approx
\langle x^{(m)}|e^{-\epsilon {\hat U_m}}|x^{(m+1)}\rangle \rho^{(m)}_0$,
where $  \rho^{(m)}_0=\langle x^{(m)}|e^{-\epsilon {\hat K}}|x^{(m+1)}\rangle$. 
The off-diagonal density matrix element involves an effective
pair interaction $U=\sum_{k,t} \Phi_{kt}$, which is approximated by its diagonal elements according to
$\Phi^{OD}_{kt}(q_{k,a},q'_{k,a},q_{t,b}, q'_{t,b};\epsilon)\approx
\frac{1}{2}[\Phi_{ab}(q_{k,a}-q_{t,b}; \epsilon)+\Phi_{ab}(q'_{k,a}-q'_{t,b};\epsilon)]$,
where the effective potential (the Kelbg potential) is given by the expression \cite{EbelForFil}:
\begin{eqnarray} 
%\begin{multline}
\Phi_{ab}(x_{ab};\epsilon) =
\frac{e_a e_b}{\tilde{\lambda}_{ab} x_{ab}} 
%\\ {} \times 
\left[1-e^{-x_{ab}^2} +
\sqrt{\pi} x_{ab} \left\{1-{\rm erf}(x_{ab})\right\} \right].
\label{kelbg-d}
\end{eqnarray} 
%\end{multline}
Here $\tilde{\lambda}_{ab}x_{ab}=|q_{k,a}-q_{t,b}|\tilde{\lambda}_e$, $\tilde{\lambda}^2_{ab}=2\pi\hbar^2\epsilon/m_{ab}$,
$1/m_{ab}=1/m_a+1/m_b$, $\tilde{\lambda}^2_{e}=2\pi\hbar^2\epsilon/m_e$ and ${\rm erf}(x)$ is the standard error function.

Note that the Kelbg potential is finite at zero distance, due to it's quantum nature.
This is a crucial advantage of the Kelbg potential over the Coulomb potential in simulations of 
systems with attractive Coulomb interaction. 
At the interparticle distance more than the thermal wavelength the Kelbg potential coincides with the Coulomb one. 
Then the off-diagonal elements of the density density matrix are:
%\begin{multline}
\begin{equation}	
\rho(\tilde{x}, \sigma ; x^\prime, \sigma^\prime ; \beta) =
\sum_{\sigma}\sum_{P_e} (\pm 1)^{\kappa_{P_e}} {\cal S}(\sigma, {\hat P_e}\sigma^\prime) 
\langle x - \xi/2| \prod_{m=0}^{M-1}
e^{-\epsilon {\hat U_{m}}} e^{-\epsilon {\hat K_{m}}}
|{\hat P_e} (x  + \xi/2)\rangle ,
\label{rho_s}
\end{equation} 
%\end{multline}
where $\tilde{x}=x-\xi/2$ and $x^\prime=x+\xi/2$ and $x\equiv x^{(0)}$. 
In the limit $M\rightarrow \infty$ the error of the whole product of high temperature factors is equal to zero $(\propto 1/M)$,
and we have an exact path integral representation of the partition function. Here 
each particle is presented by an \emph{``open''} trajectory consisting of a set of $M$ coordinates $x^{(m)}$ (``beads'').

%%%%%%%%%%%%%%%%%%%%%%%%%%%%%%%%%%%%%%%%%%%%%%%%%%%%  
It is important that these matrix elements of the density matrix can be rewritten in the form of the path integral over
\emph{``closed''} trajectories starting and ending at zero  ($ \eta^{(0)}=\eta^{(M)}=\bf 0$) \cite{filinov2020uniform}:  
\begin{widetext}
\begin{multline}
\label{pathint_wignerfunctionint4}
\rho(x, \sigma ; x^\prime, \sigma^\prime ; \beta) \big|_{x=x^\prime, \sigma'=\sigma}\,   \approx \,
\int  \rm  d\eta^{(1)} \dots \rm  d\eta^{(M-1)}
\exp\Biggl\{
-\sum\limits_{m = 1}^{M-1}
\biggl[ \pi | \eta^{(m)}|^2   +
\epsilon U\biggl(x + \sum_{k'=0}^{m}\eta^{(k')}
\biggr) \biggr]   \Biggr\}
{\rm det}||\psi(x)||,
\end{multline} 
\end{widetext} 
and we define 
\begin{eqnarray}
\label{psi}
||\psi(x)|| &=& \left \|e^{-{\pi} \left|r_{kt}/M\right|^2}\right\|_{\frac{N_e}{2}} \times
\left \|e^{-{\pi} \left|\tilde{r}_{kt}/M\right|^2}\right\|_{\frac{N_e}{2}}.
\end{eqnarray}
Here we introduce the following notation for the dimensionless coordinate  
$  
r_{kt}\equiv (x_{k,e}-x_{t,e}), \quad   (k,t=1,\dots,N_e/2),  
\tilde{r}_{kt}\equiv (x_{k,e}-x_{t,e}), \quad  (k,t=N_e/2+1,\dots,N_e) 
$
and assume that in the thermodynamic limit the main contribution to the sum over spin variables comes from the term related
to the equal number ($N_e/2$) of electrons with the same spin projection. 

In order to calculate thermodynamic functions in the canonical ensemble, the logarithm of the partition function has to be differentiated by variables $T$ and $V$.
For example, for pressure and internal energy:
\begin{eqnarray}
\beta p &=& \frac{\partial {\rm ln} Z}{\partial V} = \left[\frac{\alpha}{3V}
\frac{\partial{\rm ln} Z}{\partial \alpha}\right]_{\alpha=1}, \label{p_gen}
\\
\beta E &=& -\beta \frac{\partial {\rm ln} Z}{\partial \beta},
\label{e_gen}
\end{eqnarray}
where $\alpha= L/L_0$ is a length scaling parameter.
For pressure we have: 
\begin{eqnarray} 
\nonumber\\&&
\frac{\beta p V}{N_e+N_p}=1 - \frac{(3Z)^{-1}}{N_e+N_p}   
%\nonumber\\&&
\times \int dx d\eta^{(1)} \dots \rm  d\eta^{(M-1)} \, 
\nonumber\\&&
\times \exp\Biggl\{ -\sum\limits_{m = 0}^{M-1} \biggl[ \pi | \eta^{(m)}|^2   + 
\epsilon U\biggl(x + \sum_{k'=0}^{m}\eta^{(k')}
\biggr) \biggr]   \Biggr\} {{\rm det} ||\psi||}
%\rho(q^{(0)}, \eta^{(1)} \dots \eta^{(M)}, \beta) 
\nonumber\\
&&\times \Bigg\{ \sum_{k=1}^{N_p}\sum_{t=1}^{N_e} |x_{kt}|
\frac{\partial \epsilon\Phi_{ep}}{\partial |x_{kt}|}
+\sum_{k<t}^{N_p} |x_{kt}| \frac{\partial \epsilon\Phi_{pp}}{\partial |x_{pt}|}
+ \sum_{k<t}^{N_e} |x_{kt}| \frac{\partial
	\epsilon\Phi_{ee}}{\partial |x_{kt}|} 
\nonumber\\&&
+\sum_{m=0}^{M-1} \Big( %\left[ 
\sum_{k=1}^{N_p}\sum_{t=1}^{N_e}B(x^m_{kt})
\frac{\partial \epsilon\Phi_{ep}}{\partial |x^m_{kt}|} 
%\nonumber\\
%&&
+\sum_{k<t}^{N_e} A(x^m_{kt}) \frac{\partial 
	\epsilon\Phi_{ee}}{\partial |x^m_{kt}|} 
%\Big) %\right]
%\nonumber\\&& 
 + \sum_{k<t}^{N_p} A(x^m_{kt}) \frac{\partial \epsilon\Phi_{pp}}{\partial |x^m_{kt}|} 
\Big) %\right]
\nonumber\\&&-
\frac{\alpha}{{\rm det} ||\psi||}
\frac{\partial{\rm \,det} || \psi ||}{\partial
	\alpha}  \Bigg\} |_{\alpha=1},
\label{eeos}
\end{eqnarray}
where $B(x^m_{kt}) = \frac{\langle 	x^m_{pt}|x_{kt}\rangle}{|x^m_{pt}|}$,  $A(x^m_{kt}) =
\frac{\langle x^m_{kt}|x_{kt}\rangle}{|x^m_{kt}|}$. 

\subsection{Reducing the finite size effects} 
Computer simulations of disordered systems, such as plasmas, require an accurate accounting of the long--range Coulomb
forces. To reduce the finite--size effects in the PIMC simulations periodic boundary conditions (PBC) 
are usually imposed on the main Monte Carlo cell and the Ewald summation method is used to calculate the contribution of 
its periodic images. 
For high electron degeneracy the thermal wavelength $n\lambda^3$ can exceed the
main MC cell size $L$.  
Therefore the trajectories of electrons in the main cell can penetrate into periodic neighboring images of the main cell.
This issue requires modified PBC in the treatment of the exchange interaction in PIMC simulations.
The modification of PBC in this case is discussed in \cite{EbelForFil}. 
In addition, it is necessary to take into account the exchange interaction between particles in the main MC cell and their periodic images.
    
%%%%%%%%%%%%%%%%%%%%%%%%%%%%%%%%%%%%%%%%%%%%%%%%%%%%%ESTIMators  
To take into account the exchange and long--range Coulomb interaction between particles of the main MC cell 
and its periodic images, we can redefine the related MC estimators. The corresponding technique and the energy estimator can be found elsewhere \cite{filinov2020uniform}.
For pressure the estimator replace Eq.~(\ref{eeos}) has the following modified form:
\begin{eqnarray} 
%\nonumber\\&&  
\frac{\beta p V}{N_e+N_p} &=& 1 - \frac{(3Z)^{-1}}{\, N_e+N_p } \times  
\nonumber\\&&
\times \int dx \mathrm{d\eta^{(1)} \dots d\eta^{(M-1)}} \, 
\exp\Biggl\{ -\sum\limits_{m = 0}^{M-1} \biggl[ \pi | \eta^{(m)}|^2   + 
\epsilon \tilde{U}\biggl(x + \sum_{k'=0}^{m}\eta^{(k')}
\biggr) \biggr]   \Biggr\}  {\rm det}||\psi(x)|| 
%\rho(q^{(0)}, \eta^{(1)} \dots \eta^{(M)}, \beta) 
%\nonumber\\
\nonumber\\&&
\times \Bigg\{ \sum_{\boldsymbol{n}} \Bigg ( \sum_{k=1}^{N_p}\sum_{t=1}^{N_e} |x_{kt}|
\frac{\partial \epsilon\Phi_{ep}\Big(\boldsymbol{n}L + x  \Big)}{\partial |x_{kt}|}  
+\sum_{k<t}^{N_p} |x_{kt}| \frac{\partial \epsilon\Phi_{pp}\Big(\boldsymbol{n}L + x  \Big)}{\partial |x_{pt}|}
%\nonumber\\&&
+ \sum_{k<t}^{N_e} |x_{kt}| \frac{\partial
	\epsilon\Phi_{ee}\Big(\boldsymbol{n}L + x \Big) }{\partial |x_{kt}|} \Bigg )  
\nonumber\\&&
+\sum_{m=0}^{M-1} \sum_{\boldsymbol{n}} \Bigg( %\left[ 
\sum_{k=1}^{N_p}\sum_{t=1}^{N_e}B(x^m_{kt})
\frac{\partial \epsilon\Phi_{ep}\Big((\boldsymbol{n}L + x + \sum_{k'=0}^{m}\eta^{(k')})^m_{kt} \Big)}{\partial |x^m_{kt}|}+
\nonumber\\&&
%&&
+\sum_{k<t}^{N_e} A(x^m_{kt} ) \frac{\partial 
	\epsilon \Phi_{ee}(\boldsymbol{n}L + x + \sum_{k'=0}^{m}\eta^{(k')})^m_{kt}}{\partial |x^m_{kt}|} 
\nonumber\\
&& + \sum_{k<t}^{N_p} A(x^m_{kt}) \frac{\partial
	\epsilon\Phi_{pp}(\boldsymbol{n}L + x + \sum_{k'=0}^{m}\eta^{(k')})^m_{kt} }{\partial |x^m_{kt}|} 
\Bigg ) %\right]
\nonumber\\&&
%&&-
- \frac{\alpha}{{\rm  det} ||\psi(x)||}
\frac{\partial \sum_{\boldsymbol{n}} {\rm \,det} || \psi(\boldsymbol{n}L+x) ||}{\partial \alpha }  |_{\alpha=1} 
\Bigg\},
\label{eos}
\end{eqnarray}
where  $\tilde{U} =U_{ee}$ as the positive charges simulating neutralizing  rigid background have to be uncorrelated.  

We also use a modified Ewald scheme proposed by Yakub~\cite{yakub2003efficient} to eliminate artificial non-isotropic effects caused by PBC. The generalization of this scheme using the angle averages for electrostatic (via the Kelbg potential) and exchange interactions \cite{filinov2020uniform}.

\subsection{Basic expressions for the entropy change}
During an expansion of a system of particles both temperature and volume may change. Then 
the total entropy change $dS(T,V)$ is: 
\begin{equation}\label{ds}
  dS(T,V)=\left( \frac{\partial S}{\partial T}\right)_{V} dT  +\left(\frac{\partial S}{\partial V}\right)_{T} dV. 
\end{equation} 
To calculate the entropy change between two states $S(T_2,V_2)-S(T_1,V_1)$ we can do integration first 
at constant  $V$ and then at constant $T$ 
\begin{equation}\label{ds} 
S(T_2,V_2)-S(T_1,V_2)=\int  _{T_1}^{T_2}\left.\left( \frac{\partial S}{\partial T}\right)\right|_{V = V_2}  dT,  \quad 
S(T_1,V_2)-S(T_1,V_1)=\int  _{V_1}^{V_2}\left.\left( \frac{\partial S}{\partial V}\right)\right|_{T = T_1}  dV.   
\end{equation} 

Here as an application of the developed approach we are going to calculate the entropy change 
%at constant $\rm V$ and constant $T$ 
by integrating over $T$ and  $V$:
\begin{eqnarray} \label{ds1}
\nonumber\\&&
\Delta S_{V_2}=S(T_2,V_2)-S(T_1,V_2) = \int  _{T_1}^{T_2}\left.\left( \frac{\partial S}{\partial T}\right)\right|_{V = V_2} dT = 
\int  _{T_1}^{T_2}\left.\left( \frac{C_V}{T} \right)\right|_{V = V_2} dT = 
%\nonumber\\&&
\int  _{T_1}^{T_2}\left.\left( \frac{1}{T} \frac{\partial E}{\partial T} \right)\right|_{V = V_2} dT, 
\label{ds}
%\nonumber
\\&& 
\Delta S_{T_1}=S(T_1,V_2)-S(T_1,V_1)=\int  _{V_1}^{V_2}\left.\left( \frac{\partial S}{\partial V}\right)\right|_{T = T_1}  dV  = 
\int  _{V_1}^{V_2} \left.\left( \frac{\partial P(T,V)}{\partial T}\right)\right|_{V} dV 
%\nonumber
\label{ds1}
\end{eqnarray} 
where we use the Maxwell relation for the derivatives of entropy and pressure based on the equalities of the second derivatives of Helmholtz free energy $F$:  
\begin{equation}\label{sv}
\left(\frac{\partial S}{\partial V}\right)_T =-\frac{\partial^2 F(T,V)}{\partial V \partial T} =-\frac{\partial^2 F(T,V)}{\partial T \partial V} = 
\left(\frac{\partial P}{\partial T}\right)_V. 
\end{equation} 

\section{Simulation results}\label{simulations}

\subsection{Pair distribution functions}
%Here  we present  the results of FPIMC simulations of the unpolarized UEG.
To calculate UEG pressure and energy we use the estimators 
% estimators presented by the Eqs.~(\ref{eos}),(\ref{energy}) 
with the PBC modifications discussed above. 
We have considered electron density defined by the Brueckner parameter $r_s = \langle r \rangle / a_B$ in the range  $0.7\le r_s \le 1$ and temperature of the system in the range
$1.73\le T/\text{Ry} \le 2$, where $\langle r\rangle$ is the mean distance between electrons, $a_B$ is the Bohr radius,
Ry is hydrogen ionization energy. 

Let us start from the physical analysis of the spatial arrangement of electrons and positive particles,  
simulating the rigid neutralizing background, by studying the radial distribution function (RDF) $g_{ab}(R)$ defined as: 
%%----------------------------------------------------------------------------------------------------------------
\begin{eqnarray}\label{g-def} 
g_{ab}(|{\bf R}_1-{\bf R}_2|) &=&
%\nonumber\\
%\frac{1}{{\widetilde Z}}
\left(\frac{V}{N}\right)^2 %\left(N,V,\beta\right) \text{,}}
\sum_{\sigma}
\sum_{i,j,i\neq j}
\delta_{a_i,a}\, \delta_{a_j,b}\frac{1}{Z}
\int \rm dr\; \delta({\bf R}_1-{\bf r}_i)\,\delta({\bf R}_2-{\bf r}_j)\;
%\rho(r,Q, \sigma; %N_u,N_d,N_s,\overline{N}_{u},\overline{N}_{d},\overline{N}_{s},N_g; \beta), \{N\}; \beta),
\,\rho(x, \sigma ;\beta), 
\end{eqnarray}
where $a_i$ and $b_j$ denote the types of particles. 
%The RDF gives the probability density to find a pair of particles of types
%$a$ and $b$ at a certain distance $R=|{\bf R}_1-{\bf R}_2|$ from each other.
The RDF depends only on the difference of coordinates $R=|{\bf R}_1-{\bf R}_2|$ because of
the translational invariance of the system.
In a non-interacting classical system $g_{ab}(R)\equiv 1$, whereas interparticle interactions and  quantum statistics result in
a redistribution of particles. The product $R^2 g_{ab}(R)$ is proportional (up to a constant factor) to
the probability to find a pair of particles at a distance $R$ from each other.

\begin{figure*}[htp]
	\includegraphics[width=8.1cm,clip=true]{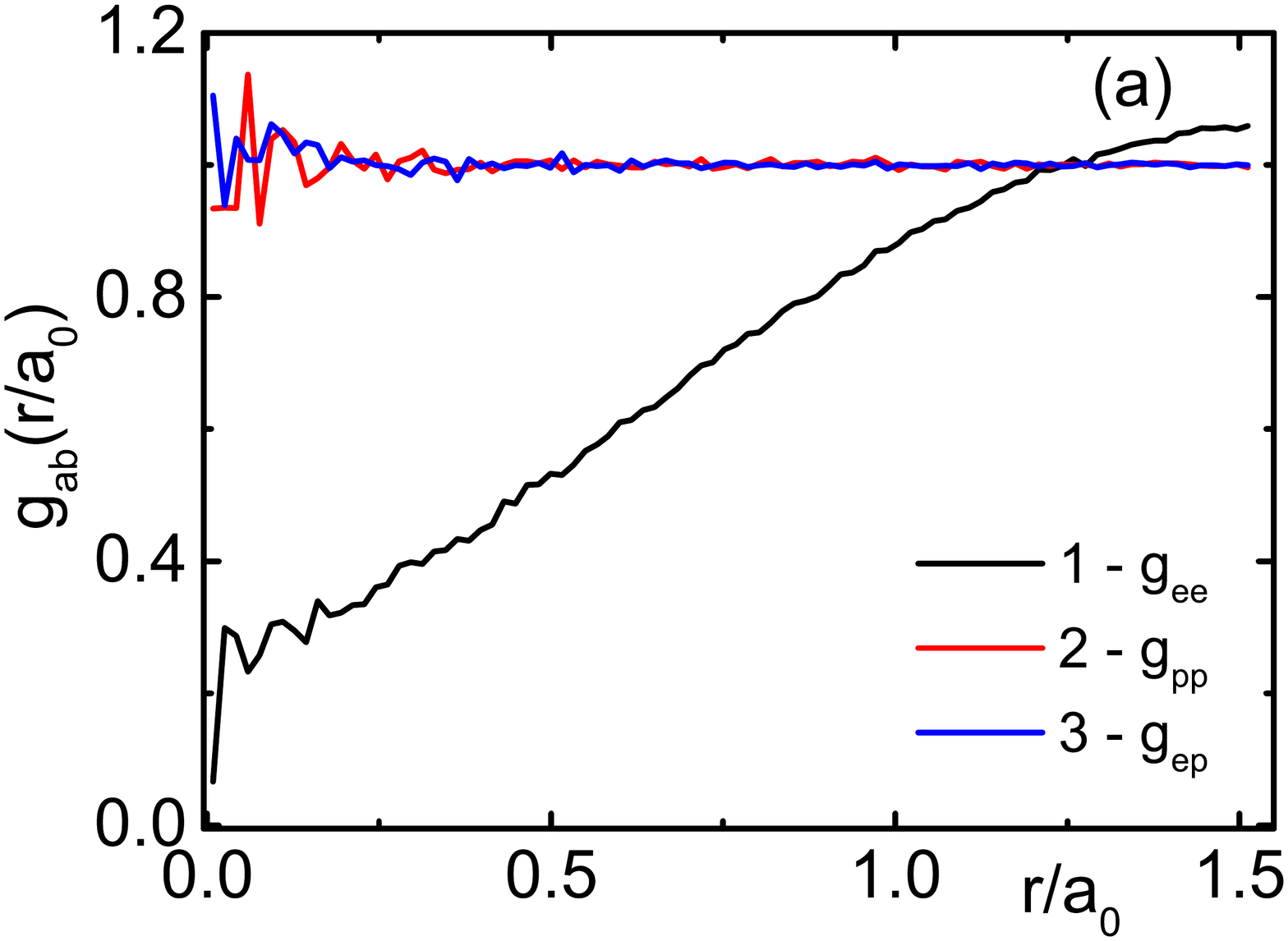}	
	\includegraphics[width=8.10cm,clip=true]{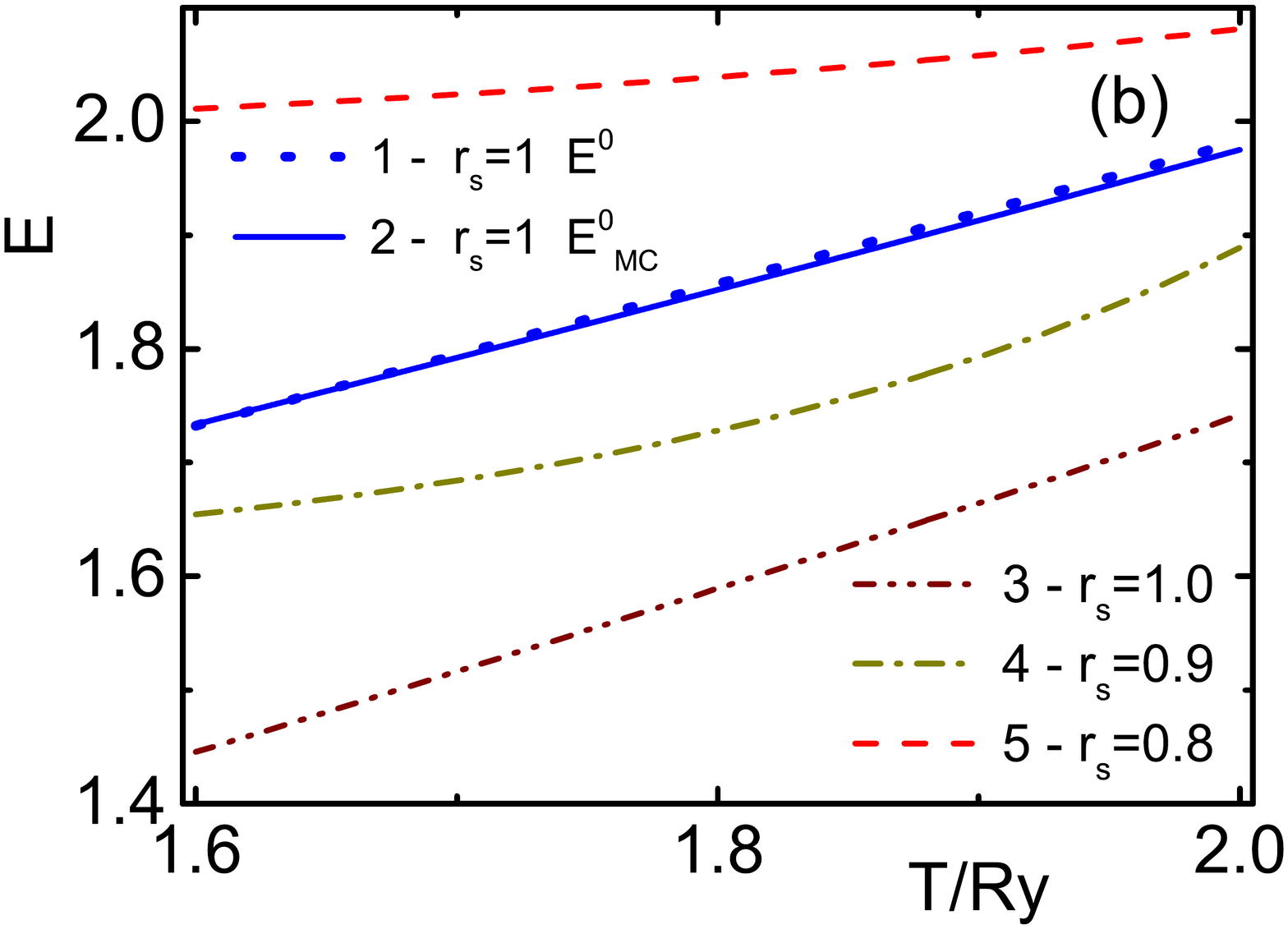}
	%\hspace{.2cm}
	\includegraphics[width=8.10cm,clip=true]{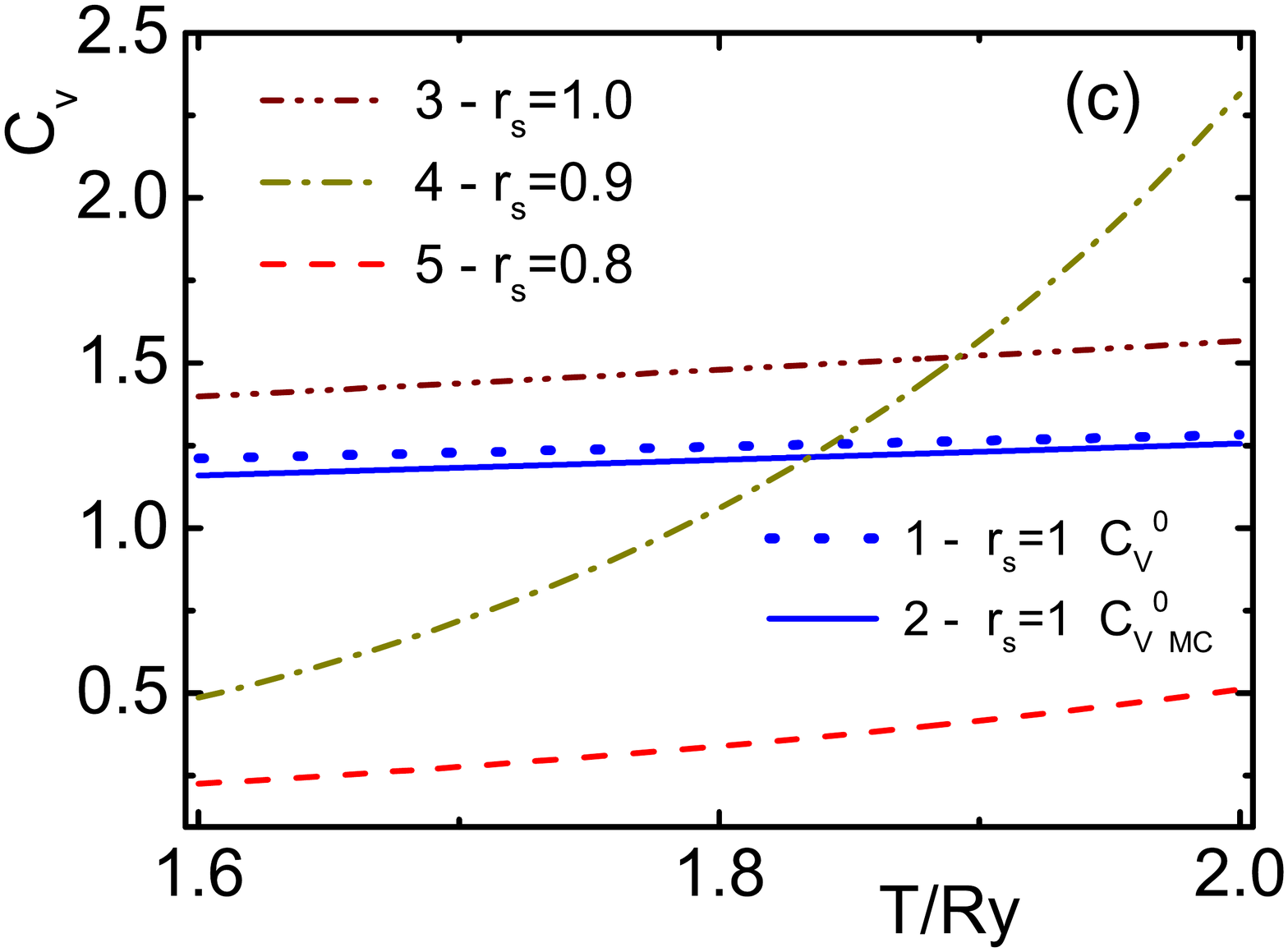}
	\includegraphics[width=8.10cm,clip=true]{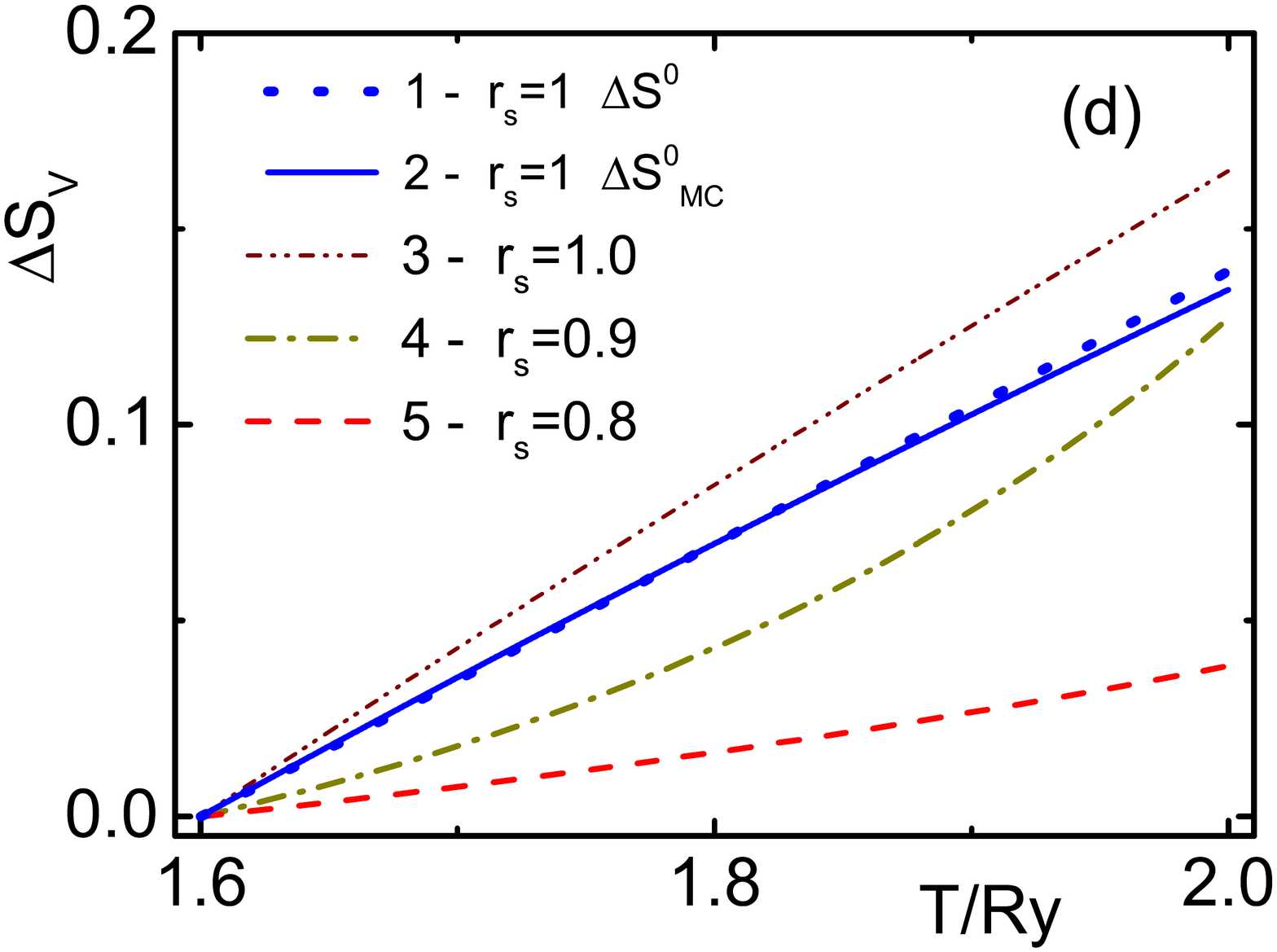}
	%	\includegraphics[width=8.1cm,clip=true]{gmet05.eps}
	%[0.4cm]
	%	\includegraphics[width=8.1cm,clip=true]{gt125.eps}
	%	\includegraphics[width=8.1cm,clip=true]{gmet25.eps}
	%\hspace{.2cm}
	%\includegraphics[width=8.1cm,clip=true]{GEtotT625.eps}
	%\includegraphics[width=8.1cm,clip=true]{PPT625.eps}
	%\hspace{.2cm}  
\caption{(Color online) (a) Electron---electron, positive charge---positive charge and 
	electron---positive charge RDFs at temperatures $T/\text{Ry} =2$ and $r_s=0.8$; 
	(b) isochores of internal energy; (c) heat capacity at constant volume; 
	(d) entropy change of UEG. 
	% in the temperature interval from $1.6Ry$ up to $2Ry$. 
	Line 1: analytical dependences for ideal UEG at $r_s=1$.
	The FPIMC simulation results for ideal UEG at $r_s=1$ - line 2; 
	for strongly coupled UEG at: %	$r_s=0.7$---line 2; 
   $r_s=0.8$---line 3; $r_s=0.9$---line 4; $r_s=1$---line 5. 
	%, $r_s=$ - line 6.   
\label{es} 
}
\end{figure*}

As an example different RDFs averaged over spin of electrons are shown in Fig.~\ref{es}a 
for a temperature $T = 2 \text{Ry}$ and density related to $r_s=0.8$. 
The positive particles are uncorrelated as they are simulating the positive rigid background in UEG and 
$g_{pp}$ and $g_{ep}$ RDFs are identically equal to unity as in ideal  gas. 

The electron-electron RDF $g_{ee}$ demonstrates a drastic difference at short distances 
due to the Coulomb and Fermi repulsion. Here $g_{ee}$ decreases monotonically 
when the interparticle distance goes to zero  but tends to unity in the opposite case. 
Oscillations of the RDF at very small distances are related to Monte-Carlo statistical error,
as the probability to find particles at short distances quickly decreases.

\subsection{Entropy changes on isochores and isotherms} 

Plots (b), (c) and (d) in Figure~\ref{es} present internal energy, isochoric heat capacity and entropy change on different isochores
for UEG in the temperature range $1.6\text{Ry}$--$2\text{Ry}$. Comparison with ideal UEG 
in Figure~\ref{es}(b) shows that the interparticle interaction downgrades the internal energy at $r_s \sim 1-0.9$ 
(degeneracy parameter $n\lambda^3\sim 3-5$ and coupling parameter $\Gamma \sim 1 $), 
but then raises it with the growth of electron density ($r_s \sim 0.8$, degeneracy parameter $n\lambda^3\sim 8$ and
practically the same coupling parameter $\Gamma \sim 1-2$).

In Figure~\ref{es}(c) heat capacity equal to the derivative of the internal energy with respect to 
temperature demonstrates more complicated behavior at temperatures less and more than $T/\text{Ry}\sim 1.8$ that can be 
the consequence of the interplay of the electron interaction and Fermi repulsion. 
Stronger electron degeneracy at $r_s\lesssim 0.9$ and $T/\text{Ry} \lesssim  1.8$ results 
in the heat capacity decrease (lines 4 and 5) in comparison with the ideal UEG.  Just on the contrary at lower density ($r_s\sim 1$)  
and $T/\text{Ry}\gtrsim1.8$ heat capacity (lines 3 and 4) is larger than that in the ideal UEG.  
%The more complicated behavior of the heat capacity in Figure~\ref{es}(c) can be understood from the consideration 
%of the temperature dependences of the internal energy in Figure~\ref{es}(b) as heat capacity 
%is the derivative of internal energy with respect to temperature. 

Entropy change in Figure~\ref{es}(d) more or less repeats the behavior of heat capacity as entropy is defined by the antiderivative of the ratio of heat capacity to temperature according  to Eq.~\ref{ds}. 
%The entropy change in Figure~\ref{es}(d) to a large extent repeats the trends 
%in behavior of the heat capacity as it is defined by the derivative of the ratio of the heat capacity to temperature according 
%to Eq.~(\ref{ds}). 

\begin{figure*}[htp]
	\includegraphics[width=8.1cm,clip=true]{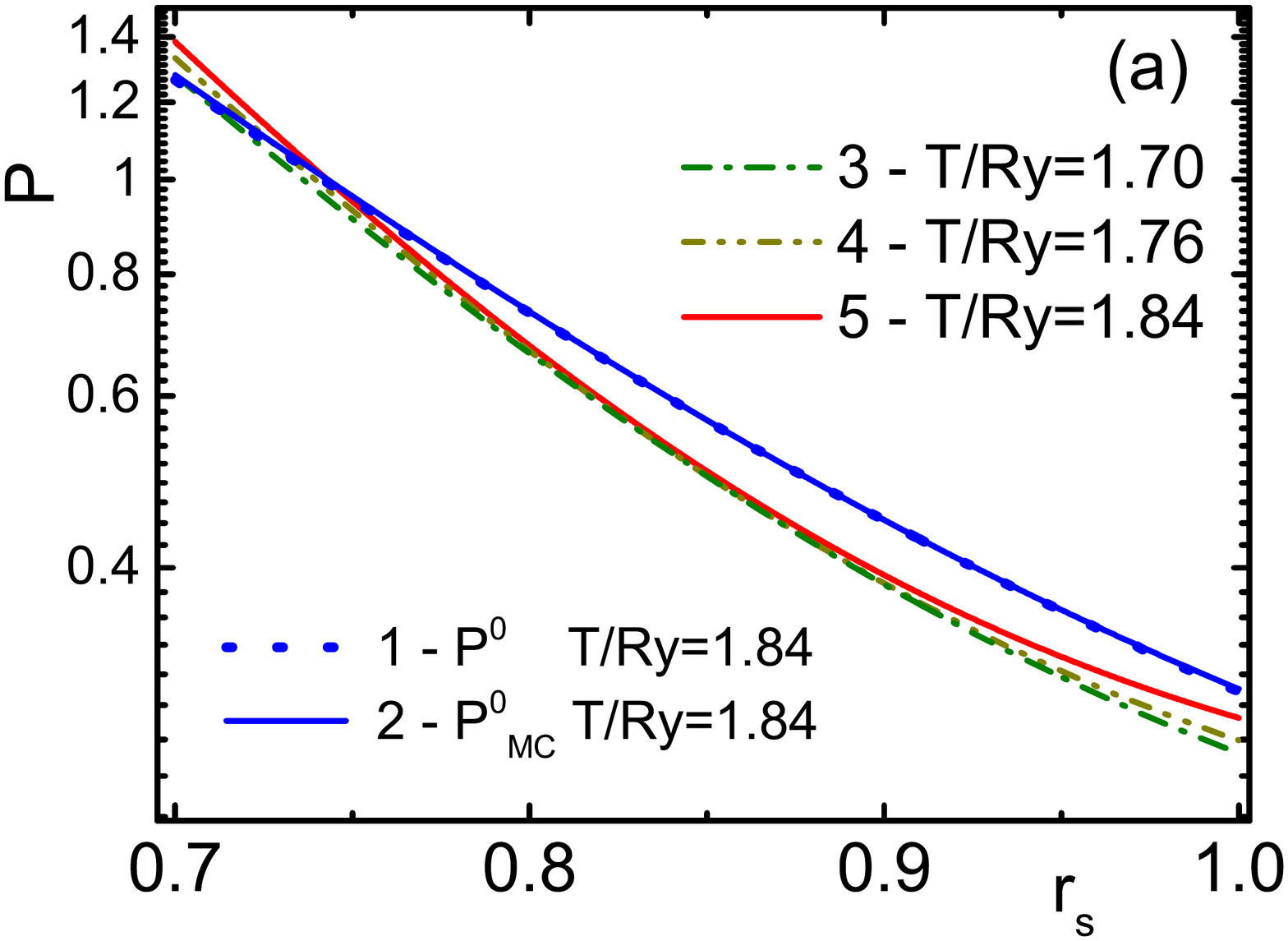}
	%\hspace{.2cm}
	\includegraphics[width=8.1cm,clip=true]{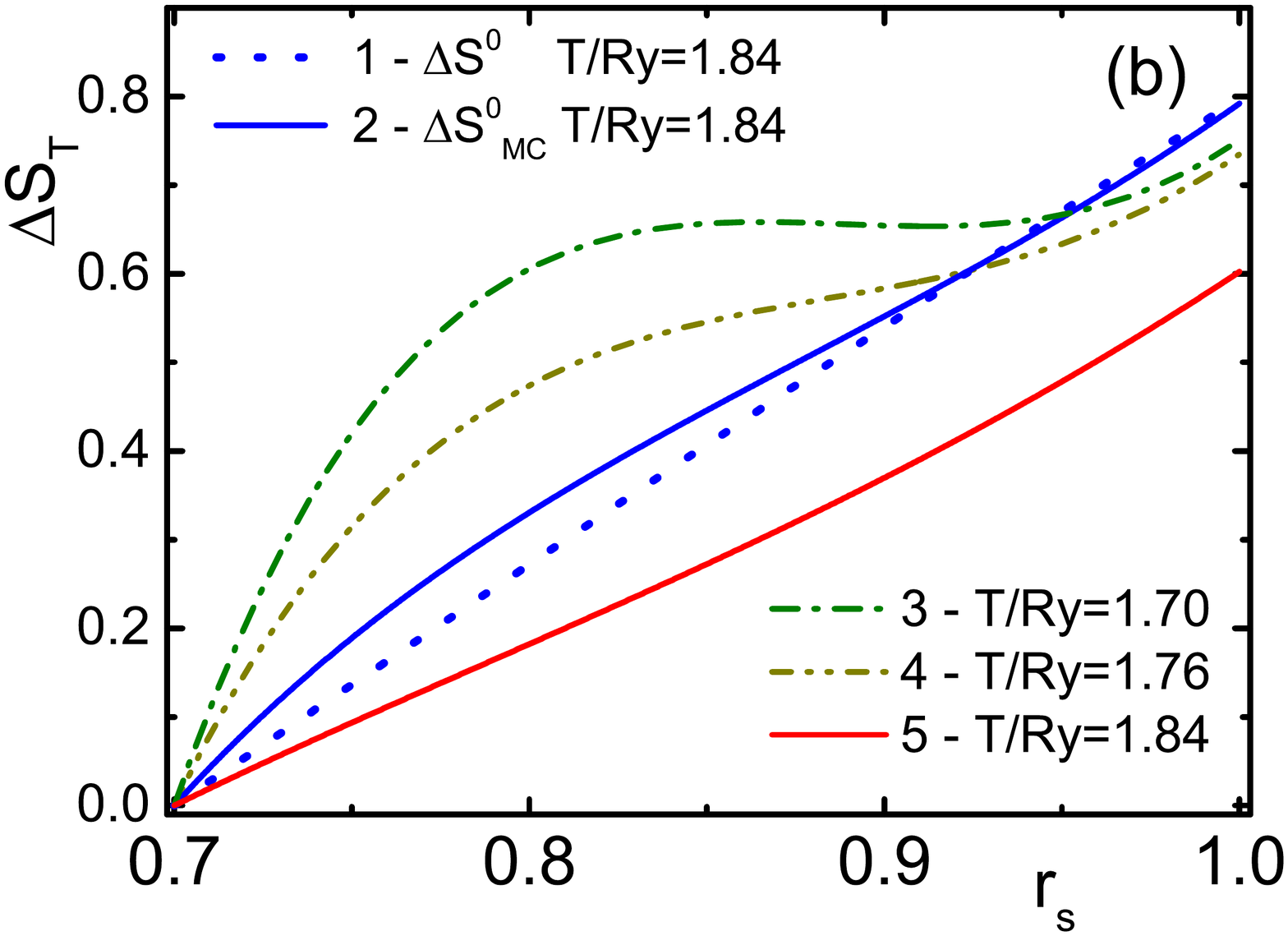}
	%	\\\includegraphics[width=8.1cm,clip=true]{gt25.eps}
	%	\includegraphics[width=8.1cm,clip=true]{gmet05.eps}
	%[0.4cm]
	%	\includegraphics[width=8.1cm,clip=true]{gt125.eps}
	%	\includegraphics[width=8.1cm,clip=true]{gmet25.eps}
	%\hspace{.2cm}
	%\includegraphics[width=8.1cm,clip=true]{GEtotT625.eps}
	%\includegraphics[width=8.1cm,clip=true]{PPT625.eps}
	%\hspace{.2cm}
	\caption{(Color online)
		Pressure (a) and entropy change (b) of UEG on isotherms. 
		% ($\theta \approx 0.2 \div 0.5$ and $\Gamma \approx 1 \div 2 $).  
		Analytical dependences for ideal UEG at $T/Ry=1.84$ are shown by lines 1. The FPIMC simulation results are: ideal UEG 
		at $T/Ry=1.84$---line 2; strongly coupled UFG at $T/Ry=1.70$---line 3; at $T/Ry=1.76$---line 4; at $T/Ry=1.84$---line 5. 
%		; $\rm T/Ry=1.84$ - line 6.  
		\label{ps}
	}
\end{figure*}

Figure~\ref{ps} shows pressure and entropy change of UEG on isotherms in the range of $r_s$ 0.7--1. 
Isotherms for pressure and entropy change of the strongly coupled UEG are presented by lines 3--5. 
Particle interaction in UEG results in decreasing pressure in comparison with the ideal UEG at low density ($r_s \ge 0.75 $). 
The opposite situation is at higher density ($r_s \le 0.75 $), where the Fermi repulsion is more important. 
From the analysis of pressure in Figure~\ref{ps}(a) 
we can expect that the derivate of pressure with respect to  temperature is a decreasing function of $r_s$ 
%as the ratio of the change in pressure to the change in temperature at a fixed density ($r_s=const$)  
as the typical difference in pressure at fixed $r_s$  on the left hand 
side of Figure~\ref{ps}(a) is larger than the same difference  on the right hand side of 
the same plot, while in between this difference is approaching zero. 
Consequently, according to Eq.~\ref{ds1} the change in entropy equals to the antiderivative over 
volume of the function $ \frac{\partial P(T,V)}{\partial T}\vert_{T}$  
can monotonically or non-monotonically increase versus $r_s$, as shown in Figure~\ref{ps}(b).  
%\textbf{\color{red} In this subsection I have partially restore the text to exclude some misunderstanding.}
 
\subsection{Error estimations}

According to Eqs.~(\ref{ds}) and (\ref{ds1}) entropy change is defined by the integral over the derivatives 
of pressure and internal energy with respect to temperature. In the used FPIMC approach these derivatives can not be 
directly calculated and have been obtained by the numerical differentiation of the smoothed FPIMC results  
for these functions on the corresponding isotherms and isochores. 
Sure the accuracy of numerically obtained derivatives is lower than the accuracy of the same values calculated directly 
(by any other MC method). Moreover errors in calculations of entropy change can be accumulated at integration over the 
obtained approximations of derivatives according to Eqs.~(\ref{ds}) and (\ref{ds1}). 

As an example, the estimation of the accuracy of the presented FPIMC results for entropy change can be obtained from 
the comparison of lines 1 and 2 presenting entropy change for ideal UEG in Figure~\ref{es}(d). % and \ref{ps}. 
Here line 1 presents the analytical dependence, while line 2 shows the FPIMC results for the ideal UEG. 
The accuracy of the used numerical differentiation is high enough for reliable analysis of the main features  
of the physical behavior of presented data. We plan to eliminate the small discrepancy in entropy change from the analytical model
in the future. 

The main source of errors in PIMC simulations is the correct accounting of exchange interaction. 
As we mentioned above the developed FPIMC demonstrates a good accuracy for the degenerate ideal UEG 
and moreover as it follows from the paper \cite{filinov2020uniform} this is also true for the strongly 
coupled degenerate UEG. 
%Comparison of the isotherms and isochores for ideal UEG showing the analytical dependence 
%with those, presenting the FIPMC result on the Figures~\ref{ps} and \ref{es} demonstrates 
%a good accuracy of accounting for of the exchange interaction.   

\section{Discussion}\label{summary} 

This contribution presents a new PIMC approach and first results for pressure and entropy difference of the degenerate and strongly coupled electron gas. Knowledge of these quantities is of great interest 
in the study of thermodynamics and can also be used for physical explanations of the experimental observations.
Advantage of the developed approach is that it works in a broad density--temperature 
range because it takes into account Coulomb and exchange interaction not only in the main MC cell  
but also in surrounding periodic images. 
Several improvements have been made. The first one is 
%the improved treatment of exchange interaction, achieved by 
a proper change of variables modifying the path integral measure and exchange determinant. 
Then we use the angle--averaged long--range Coulomb interaction via the Ewald summation, 
as proposed by Yakub et al.\cite{yakub2003efficient} and in addition we have developed the angle averaging of the exchange determinant
describing the fermionic exchange interaction between particles in the main Monte Carlo cell and   
from the nearest images. The developed FPIMC method practically does not suffer from the  ``fermionic sign problem'',  
usually arising in PIMC simulations of degenerate Fermi systems. This particular extension is of greatest interest 
for numerical simulations as it circumvents the mathematical difficulties that one 
encounters by adapting straightforward MC strategies applicable for classical systems. 

The obtained results include pair distribution functions, isochores and isotherms of pressure, internal energy  
and entropy change of the strongly coupled and degenerate UEG. We demonstrate a strong influence of interaction 
on thermodynamic properties of the UEG especially on the entropy change. 
The developed FPIMC method practically does not suffer from the  ``fermionic sign problem'',  
usually arising in PIMC simulation of degenerate Fermi systems. 
This is attributed to the improved treatment of exchange in the FPIMC approach. 
The FPIMC simulations converge very fast and require only a few hours per one run on modern computers. 
Further development of the FPIMC ideas is now  in progress. 

\section*{Acknowledgements}

We acknowledge stimulating discussions with Prof. M. Bonitz, T. Schoof, S. Groth and T. Dornheim (Kiel). 
The theoretical development of the UEG model was supported by the Russian Science Foundation, Grant No.~20-42-04421. 
The path integral Monte Carlo method and its algorithmic realization was supported by the grant in the form of 
a subsidy for a large scientific project in priority areas of scientific and technological development No. 13.1902.21.0035. 
The extensive numerical calculations of the pair distribution functions, the isochores and isotherms of pressure and internal energy  
and the entropy changes in the strongly coupled and degenerate UEG in a wide range of density and temperature 
have been carried out in the frame of the State assignment  No.~075-00892-20-00. 

%   
%This work was supported by the Russian Science Foundation, Grant No.~20-42-04421 and 
%The work is supported by the grant in the form of a subsidy for a large scientific project in priority areas 
%of scientific and technological development No. 13.1902.21.0035 and the State assignment No.~075-00892-20-00.
%\begin{thebibliography}{61}
%	
%\end{thebibliography}
%\section*{References}
%\bibliographystyle{iopart-num}
\bibliographystyle {apsrev}
%\bibliography{FilinLar}
\bibliography{ueg.bib}

\end{document}